# Lithium Niobate Surface-Acoustic-Wave Resonator for Very-High-Frequency Isolated Power Conversion

Liyang Jin, Zhiyang Cao, *Graduate Student Member, IEEE*, Zichen Xi, Jun Ji,
Yizheng Zhu, Liyan Zhu, *Member, IEEE*, Linbo Shao, *Member, IEEE*

***Abstract*— Piezoelectric devices offer a promising magnetic-free alternative for high-frequency power electronics, enabling compact energy transfer with inherent galvanic isolation. Here we demonstrate a lithium niobate surface acoustic wave (SAW) resonator that enables isolated power transfer and addresses the challenges in existing piezoelectric devices for power applications, including operation frequencies, power delivery figure of merit (FOM), power handling, and electrical isolation. Our application-optimized 40-MHz SAW resonator, with the two electrically isolated interdigital transducers (IDTs) and acoustic Bragg mirrors, achieves a 25.7 W output power at a less than 15 °C temperature rise with only passive cooling through the back mounting side, and an isolation voltage of 2.17 kV. In DC-AC power conversion, we demonstrate a peak efficiency of 93.8%, a power delivery FOM of 13.47 mW/V$^2$ and a peak chip power density of 81.7 W/cm$^3$. DC-DC conversion is also demonstrated by connecting a diode rectifier to the output IDT, despite efficiency degrading to 52.7%. Our results advance SAW devices as one of the core components for next-generation very-high-frequency, miniaturized, high-density, magnetic-free, isolated power converters.[1]**

***Index Terms*— Surface acoustic wave resonator, piezoelectric devices, isolated power conversion, very high frequency**

## I. Introduction

Power converters are increasingly operating at higher frequencies to achieve reduced physical size, greater power density, and faster transient response. In particular, the advent of gallium nitride (GaN) devices[1, 2] is enabling switching frequency into the very-high-frequency (VHF) band (30 to 300 MHz). However, at these frequencies, conventional magnetic components experience exponentially increasing core losses due to hysteresis and eddy currents.

To circumvent the physical limits of magnetics, piezoelectric devices[3-6] have emerged as rising alternatives. Piezoelectric devices can store and transfer energy in mechanical elastic waves, offering higher power density and much higher quality ($Q$) factors than their electromagnetic counterparts at VHF. Bulk acoustic wave (BAW) devices are used in DC-DC converters with switching frequencies from low frequency (30 to 300 kHz)[7-11], medium frequency (300 kHz to 3 MHz)[12-15] to high frequency band (3 to 30 MHz)[16-19]. A record of 5.7 kW/cm$^3$ mechanical power density has been reported using a thickness-extensional mode lithium niobate (LN) resonator[18]. Multi-port piezoelectric resonators have enabled isolated conversions[10] and sensing capability[20], although their operation frequencies and power densities are much lower. In addition, these BAW devices used in power conversions face practical challenges: (1) The required suspension packaging limits heat dissipation. (2) The presence of spurious modes in thickness-mode devices complicates control and reduces overall efficiency. Extensive efforts, such as engineering shapes of mechanical resonators and their electrodes[18], are made to suppress these spurious modes. (3) Simply scaling the device to VHF would result in fragile structures with thicknesses less than 100 μm, resulting in challenging fabrication and packaging.

Alternatively, we see the opportunity that specifically designed surface acoustic wave (SAW) devices can potentially address these challenges in power applications. (1) As SAWs are inherently confined near the surface, decaying exponentially into the bulk, SAW devices can be directly surface-mounted, improving structural robustness and heat dissipation. (2) SAWs can be well defined by narrow-band Bragg mirrors, leading to much fewer spurious modes within the resonant frequency band. (3) Their resonant frequency is lithographically defined by the structures (e.g., pitches of electrode), not by the thickness or dimensions of the chip die, allowing multiple devices with different frequencies on a single die.

To date, SAW devices are widely used in microwave[21, 22] and sensing applications[23], and recently explored for mechanically-isolated gate drivers [24, 25], but have not yet been well developed as an energy storage and transfer device for power conversions. Unlike microwave signal processing applications, SAW devices for power conversion require high energy transfer efficiencies (e.g., >90%), high power (e.g., >10 W) handling capabilities, high isolation (breakdown) voltages (e.g., > 1 kV), and low impedances for high power delivery figure of merit (FOM) [10], defined as $P_{out}/V_{in}^2$. Existing SAW devices do not meet these requirements and necessitate the development of power-application-specific SAW devices. In this letter, we design and demonstrate such a 40-MHz two-port SAW resonator to address these device requirements.

## II. Device Design and Characterization

Our SAW resonator [Figs. 1(a) and 1(b)] is fabricated on a 500-μm-thick 128° Y-cut bulk LN substrate. We use LN as the piezoelectric material for its low dielectric loss, large linear region, and a high operating temperature demonstrated up to

Corresponding authors: Liyan Zhu (email: liyanz@vt.edu) and Linbo Shao (email: shaolb@vt.edu).

L. Jin and Z. Cao contribute equally to this work.

L. Jin, Z. Xi, J. Ji, Y. Zhu and L. Shao are with Bradley Department of Electrical and Computer Engineering, Virginia Tech, Blacksburg, VA 24060 USA.

Z. Cao and L. Zhu are with Center for Power Electronics Systems (CPES), Virginia Tech, Blacksburg, VA 24060 USA.

This work was in part sponsored by the Air Force Office of Scientific Research (AFOSR) under Grant Number W911NF-23-1-0235. The views and conclusions contained in this document are those of the authors and should not be interpreted as representing the official policies, either expressed or implied, of AFOSR or the U.S. Government. Device fabrication was conducted at the Center for Nanophase Materials Sciences (CNMS2024-B-02643) at Oak Ridge National Laboratory, which is a DOE Office of Science User Facility.

Competing interests: Y. Z, and L. S. are co-founders and shareholders of Tern Photonics Inc. The other authors declare no competing interests.

800 °C[26]. Our device features a footprint of 1.8325 cm × 0.57 cm, and SAWs propagate in the crystal X direction. In this configuration, the SAW velocity is about 4 km/s, and thus a 40-MHz frequency corresponds to an acoustic wavelength of 100 µm. The pitch between the interdigital transducer (IDT) electrodes is 50 µm (half wavelength), leading to an air-breakdown-limited input voltage over 400 V, and the width of an individual electrode is 25 µm. Each IDT consists of 13 pairs of electrodes, leading to a low impedance of <10 Ω and thus a high $P_{out}/V_{in}^2$. The two IDTs are separated by 500 µm, leading to a >2 kV isolation voltage. The metal Bragg mirrors have the same pitch of 50 µm. The lateral dimension of the Bragg mirrors is 5.7 mm. Both the metal mirrors and IDTs are patterned by direct-write lithography (Heidelberg MLA150), electron-beam evaporation of 250-nm-thick aluminum and followed by a lift-off process. The diced chip has a dimension of 1.9 cm (long) × 1 cm (wide) × 0.05 cm (thick).

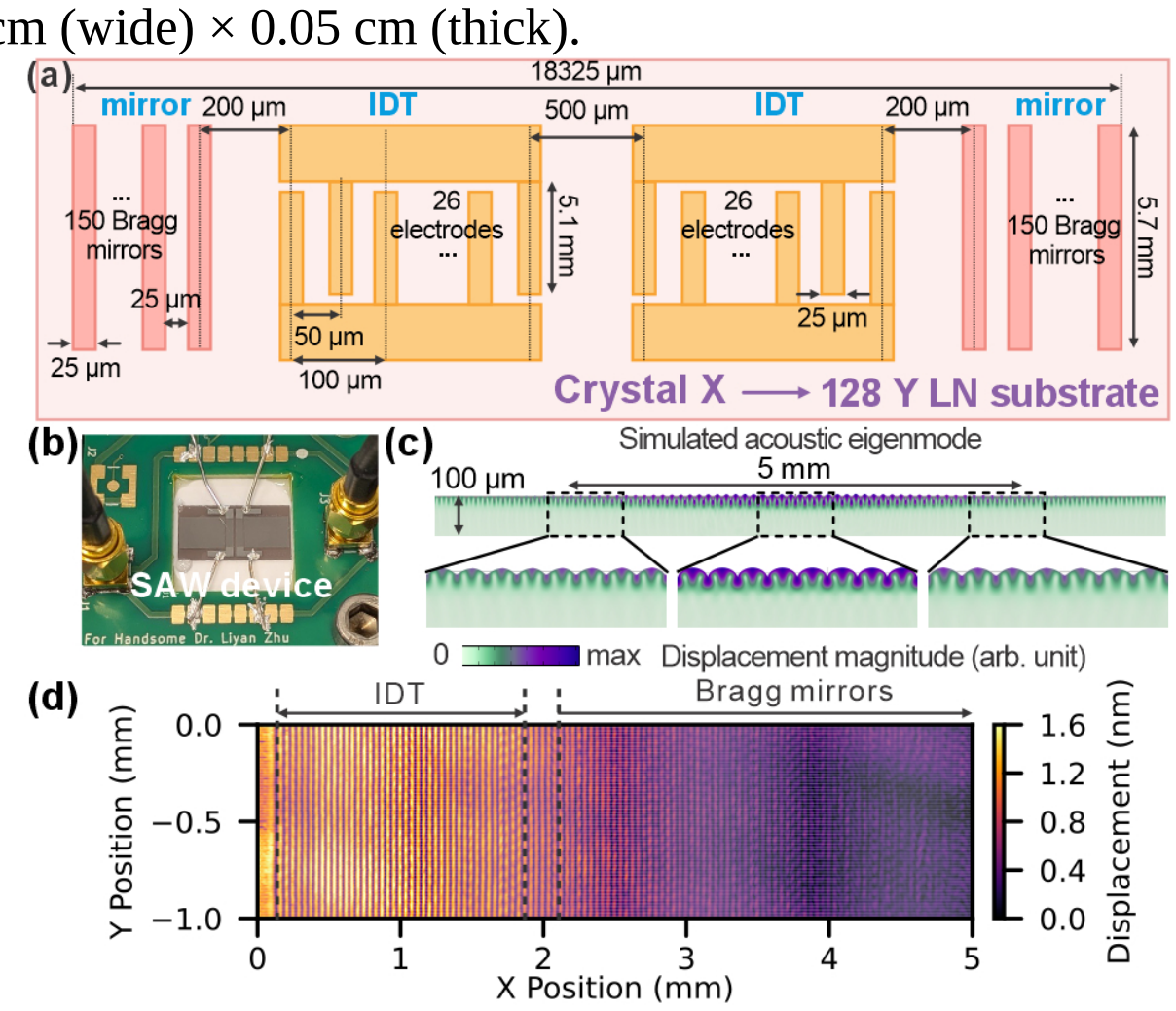

Fig. 1. (a) Design of the SAW resonator. (b) Photo of packaged SAW device; the backside of SAW chip is directly glued to a ceramic plate. (c) Numerically simulated cross section of the SAW resonant mode at 38.81 MHz, with simulation region thickness of 100 µm. (d) Measured displacement profile using an optical vibrometer from the IDT to the Bragg mirror region.

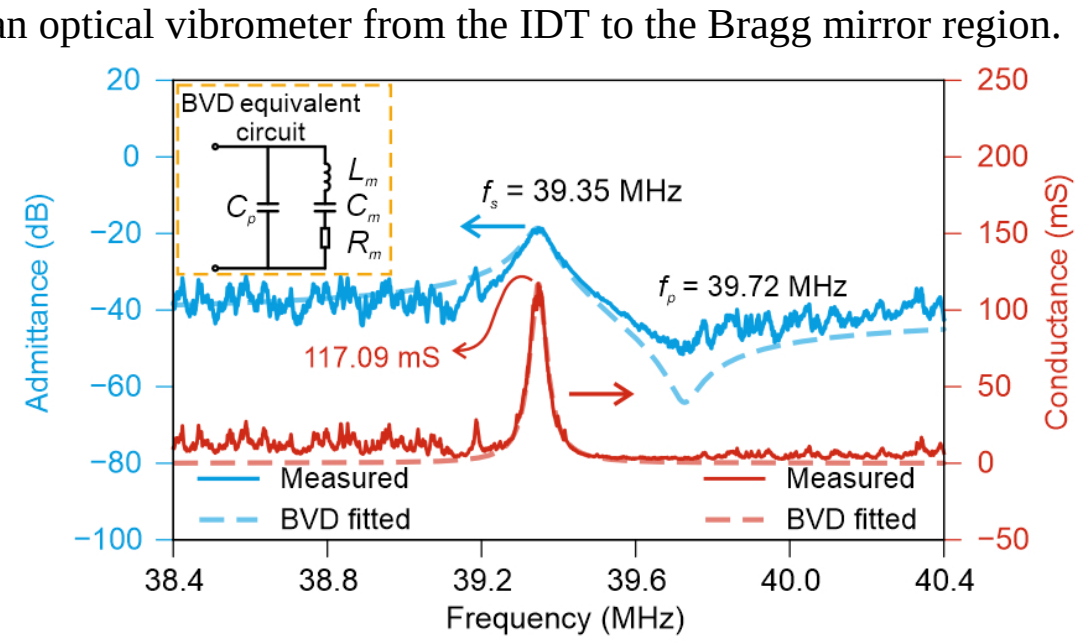

Fig. 2. Single-port Y-parameter spectra of the SAW device. The other port is open. Inset: fitted BVD equivalent circuit model to single-port measured results.

The center-to-center distance between the outmost IDT electrode and its adjacent Bragg mirror is two acoustic wavelengths (200 µm) to ensure the phase alignment between the incident and reflected SAWs, while maintaining the gap between structures. At the resonant frequency, the Rayleigh-mode SAWs are confined by the mirrors with decaying amplitude into the mirror region, as shown in the COMSOL eigen-mode simulation [Fig. 1(c)] and experimentally measured displacement profile [Fig. 1(d)][27]. We therefore estimate the active acoustic mode volume to be 1× 0.51 × 0.005 cm³. We characterize our SAW resonator using a vector network analyzer (VNA, Keysight P5000A). Figure 2 plots the admittance of the resonator in a single-IDT measurement (the other IDT is open), showing a resonance at 39.35 MHz. The measured admittance spectrum is fit to the Butterworth-Van Dyke (BVD) equivalent circuit model [28], yielding static input capacitance $C_p$ = 34.28 pF, mechanical resonance equivalent $L_m$=24.75 µH, $C_m$=0.66 pF, and $R_m$=8.54 Ω, corresponding to a BVD Q of 715. The power conversions are operating within the frequency range between the series resonant frequency $f_s$=39.35 MHz and the parallel resonant frequency $f_p$=39.72 MHz, where the piezoelectric device is inductive. As SAWs are confined by the Bragg mirrors, which do not couple SAWs into unwanted bulk modes, the operating frequency range from $f_s$ to $f_p$ remains free of significant spurious interference.

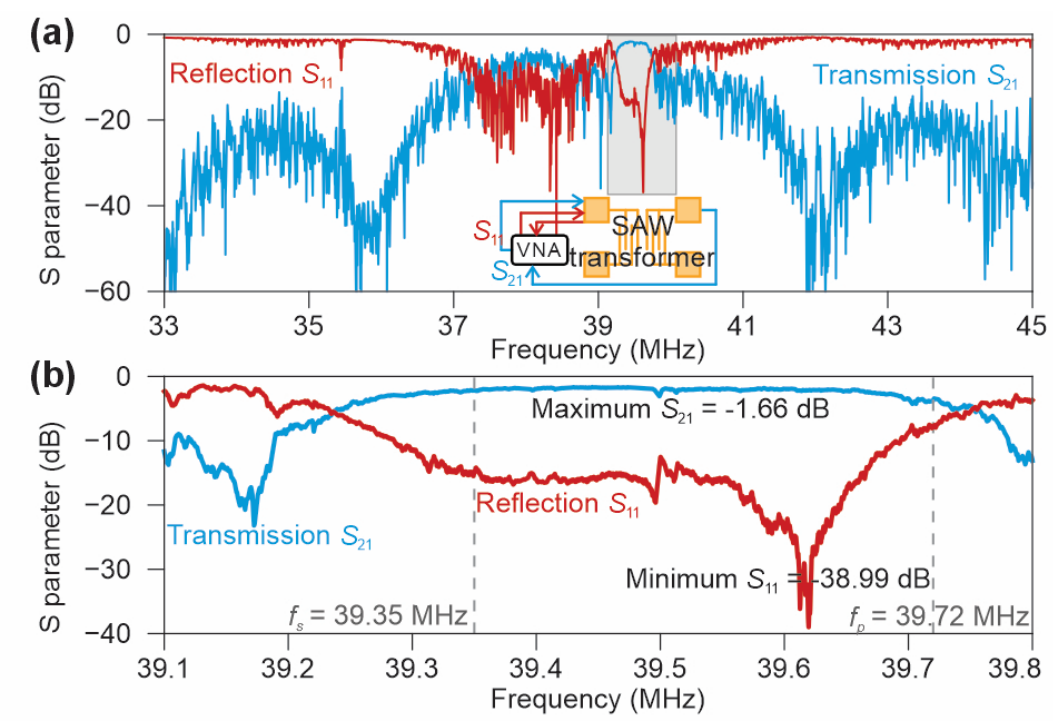

Fig. 3. Measured two-port S-parameter spectra of our SAW device. (a) Transmission and reflection S-parameter spectra. (b) Zoom-in of spectra near resonance. The characterization impedance is 50 Ω.

Figure 3 plots the two-port S-parameter spectra of our device. The sinc-function-shaped transmission $S_{21}$ pattern, with zeros at 35.66 MHz and 41.92 MHz [Fig. 3(a)], is the result of direct electromechanical transmission between IDTs without resonance enhancement. Within the frequency range between series and parallel resonant frequencies ($f_s$ and $f_p$), the electrical-to-acoustic transduction is enhanced by the acoustic resonance, and high transmission $S_{21}$ is observed with a maximum $S_{21}$ of -1.66 dB [Fig. 3(b)]. The low $S_{11}$ indicates that most electrical power is converted to surface acoustic waves at input IDT. We note that S parameters shown here are with a standard 50 Ω network and could be converted to different impedances for circuit analysis following the microwave engineering procedures[29].

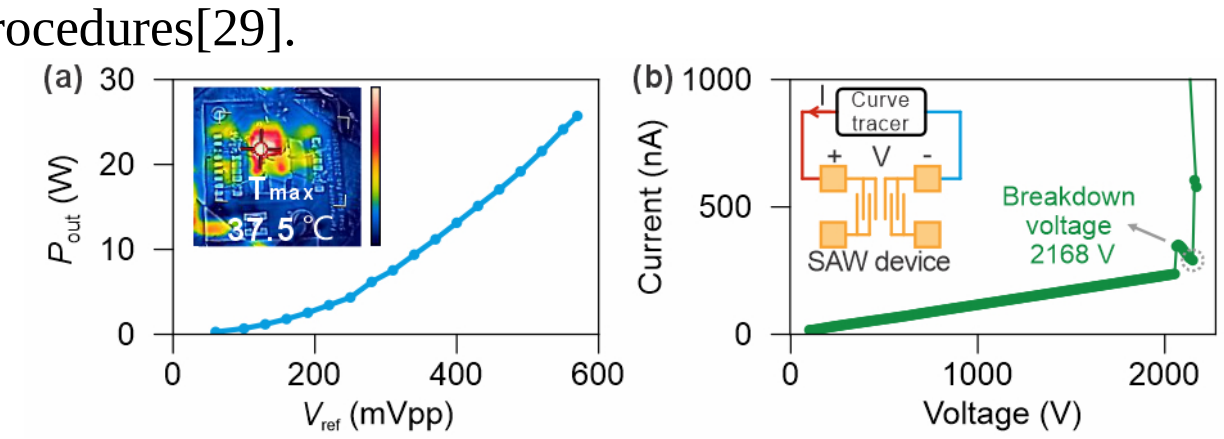

Fig. 4. (a) Measured output power of the SAW device at various reference signal amplitudes. Inset: thermal camera photo of packaged SAW device at 25 W output. (b) Measured I-V characteristic of the SAW device during breakdown measurement.

To assess the power handling, we use a power amplifier (Amplifier Research, 150A 100C) with reference signals from a function generator (Tektronix, AFG3102C) to directly drive our SAW device. While the mounted SAW device is contacted to a copper heat sink, a peak SAW device output power of 25.7 W is achieved with a 99.8 Ω load resistor, and the hot-spot temperature on the SAW resonator reaches 37.5 °C [Fig. 4(a)]. We use Keysight B1505 to measure the DC isolation (breakdown) voltage between the input and output IDTs, obtaining a value of 2.17 kV [Fig. 4(b)], limited by air breakdown between the electrodes.

## III. Demonstration of SAW-based Power Conversion

DC-AC power conversion is first demonstrated using our SAW devices, where the load resistor is directly connected to the output IDT [Fig. 5(a)]. Our SAW device is driven by a custom-built full-bridge inverter, composed of four GaN high-electron mobility transistors (Q1-Q4, EPC2037). We use two LMG1210 half-bridge gate drivers, each with a 2 ns dead time, to control the GaN full bridge. The two gate drivers are triggered by two complementary PWM signals with 50 % duty cycle, respectively. The power consumption of each half-bridge gate driver is 0.09 W. Figure 5(b) shows the measured input and output voltage waveforms. Figure 5(c) shows the peak AC output powers and efficiencies at different load resistances. The peak DC-AC conversion efficiency of 93.8% differs from VNA small-signal transmission efficiency (-1.66 dB=68.2%), as the circuit impedances are different. DC-DC conversion is also demonstrated by connecting the output IDT to a diode bridge full-wave rectifier (BAS-4004E6327HTSA1) and then a load resistor [Fig. 5(a) inset]. We perform 2D parameter sweeps of the input voltages and switching frequencies and measure the output power and power efficiencies [Fig. 6]. The reported DC-AC and DC-DC efficiencies are calculated from the measured power consumption by the load resistor divided by the measured input power to the GaN full bridge (gate river power is not included). The DC-DC conversion efficiency is significantly lower than DC-AC conversion efficiency; it is a result of diode forward voltage drop, and the varying impedance seen by the SAW resonator. Further circuit development is needed to improve DC-DC power conversion efficiency.

## IV. Discussion and Conclusion

This work presents an application-specific SAW device for miniaturized, high-density, and magnetic-free isolated power conversions. It achieves a maximum small-signal transmission of -1.66 dB at 39.35 MHz and shows a quality factor of 715. With a large AC signal input, the SAW resonator demonstrates power handling capability up to 25.7 W. We prototype DC-AC and DC-DC power converters using the SAW resonator. In DC-AC conversion, we achieve an efficiency of 93.8% at 9-V DC input, a peak output of 7.76 W at 24-V DC input, and a chip power density of 81.7 W/cm$^3$. In DC-DC conversion, a peak DC-DC efficiency of 52.7% is achieved at the 13-V to 11.9-V condition, and a peak power of 1.66 W at the 14-V to 18.2-V conversion.

Tables I and II compare our device with previous ones for DC-AC and DC-DC power conversion, respectively. In addition to the power density calculated based on the diced chip dimension (1.9 × 1 × 0.05 cm$^3$), we also provide mechanical power density based on acoustic mode volume (1× 0.51 × 0.005 cm$^3$) to show the elastic limit of LN material for future device engineering. While our SAW device achieves the highest switching frequency and highest power delivery FOM, our DC-DC conversion efficiency is significantly lower than the state of the art. Future research could focus on circuit development for SAW-based converters, optimization for specific applications (e.g., power level, output/input voltage ratio, load), and co-packaging/integration of SAW devices with GaN HEMTs.

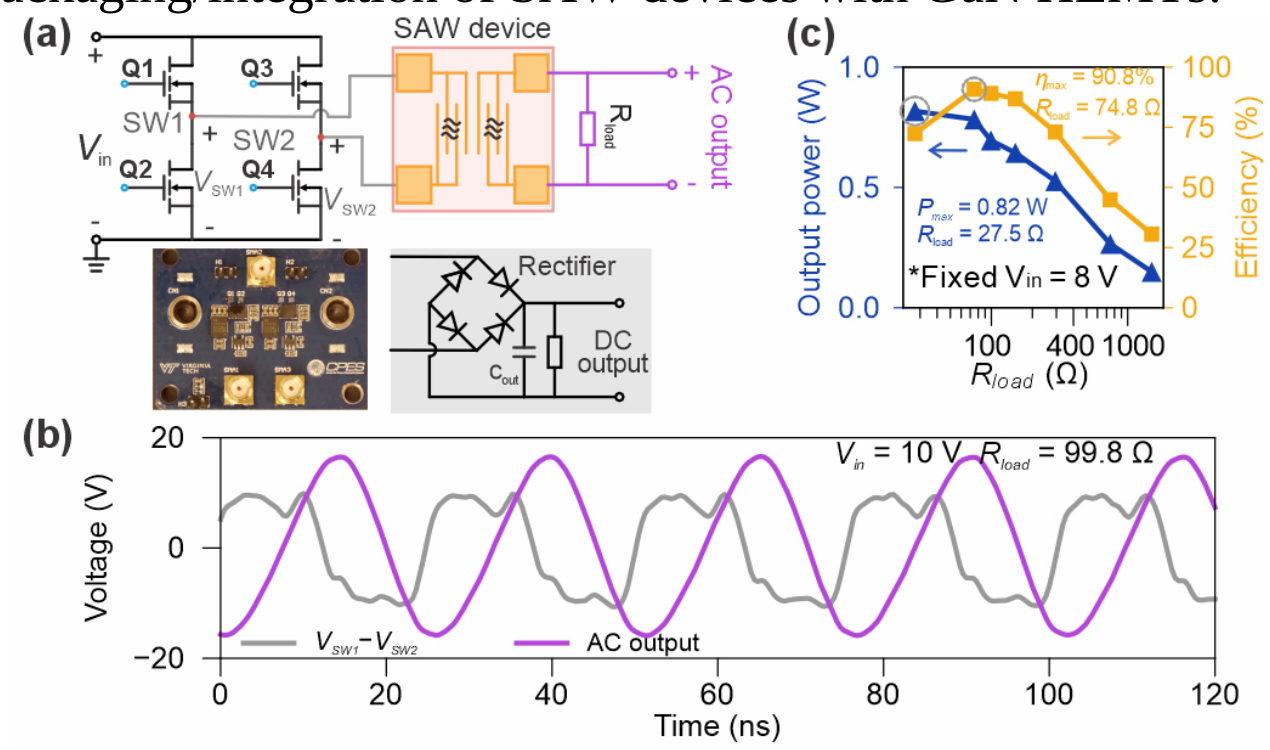

Fig. 5. (a) Setup for DC-AC and DC-DC conversion. Inset: photo of GaN full-bridge board. (b) Measured voltage waveforms of switching node $V_{SW1}$ - $V_{SW2}$ and AC output voltage at $V_{in}$ of 10 V. (c) Measured output AC power and efficiency with different load conditions at $V_{in}$ of 8 V.

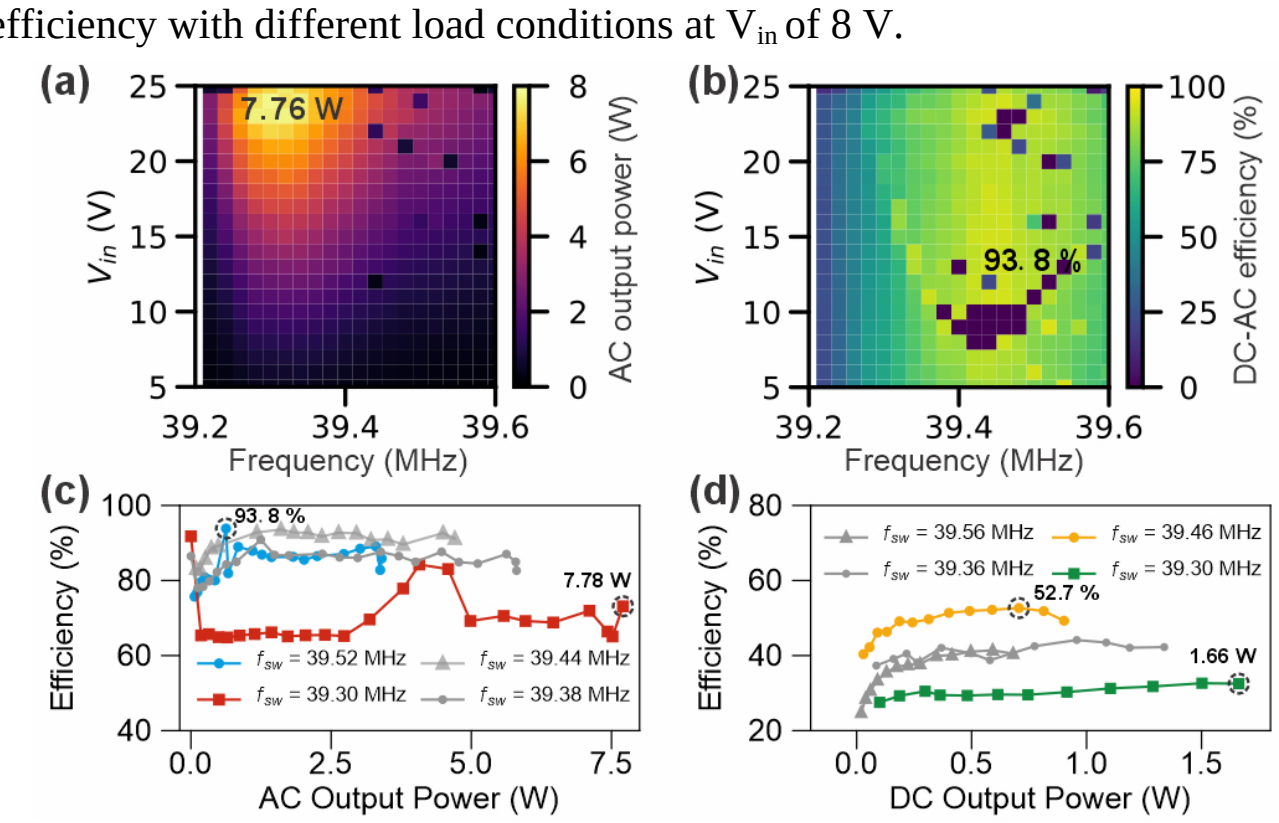

Fig. 6. Experimental characterization of DC-AC and DC-DC conversion using the SAW device. Measured DC-AC (a) output powers and (b) power-stage efficiencies at different input voltages and switching frequencies when $R_{load}$ = 99.8 Ω. (c) Measured power-stage efficiency vs. AC output power when $R_{load}$ = 99.8 Ω. (d) Measured power-stage efficiency vs. AC output power when $R_{load}$ = 200 Ω.

Table I. Comparison of Reported **DC-AC** Piezoelectric Power Devices

| Device Material-Type | Freq. (MHz) | Circuit-based $Q$ | Electrical Isolation | Peak $P_{out}$ (W) | Peak $P_{out}/V_{in}^2$ (mW/V$^2$) | Peak Power Density (W/cm$^3$) | $\eta_{max}$ (%) |
|---|---|---|---|---|---|---|---|
| PZT-Radial[7] | 0.081 | 361 | No | **32** | 0.41 | 11 | 90 |
| PZT-Rosen[30] | 0.070 | - | **Yes** | 0.5 | 0.25 | 330* | 83 |
| PZT-Radial[11] | 0.078 | **1610** | **Yes** | 10 | 0.44 | 2.36 | 65 |
| **LN-SAW (This work)** | **39.35** | 715 | **Yes (2.1 kV)** | 7.76 | **13.47** | **81.7†, 3043‡** | **93.8** |

* Theoretical value calculated at 100 kHz. † Calculated using chip dimensions of 1.9 (long) × 1 (wide) × 0.05 (thick) cm$^3$. ‡ Calculated based on acoustic mode volume of 0.51 (long)×1 (wide)×0.005 (thick) cm$^3$.

Table II. Comparison of Reported **DC-DC** Piezoelectric Power Devices

| Device Material-Type | Freq. (MHz) | Circuit-based $Q$ | Electrical Isolation | Peak $P_{out}$ (W) | Peak $P_{out}/V_{in}^2$ (mW/V$^2$) | Peak Power Density (W/cm$^3$) | $\eta_{max}$ (%) |
|---|---|---|---|---|---|---|---|
| PZT-Radial[19] | 0.095 | 1000 | No | 50 | 0.8 | 135.8 | 93.8 |
| LN-Radial[31] | 0.377 | 17000 | No | 1.22 | 0.122 | 19.42 | 88.3 |
| LN-Radial[13] | 0.75 | **27800** | No | 100 | 3.91 | 1700 | **99.3** |
| PZT-Thickness[19] | 1.04 | 500 | No | 175 | 4.38 | 176.8 | 84 |
| LN-Thickness[18] | 6.24 | 2700 | No | **1000** | 4.94 | **5700** | 97.9 |
| PZT-Radial[10] | 0.090 | 1413 | **Yes** | 4.93 | 0.5 | 18.71 | 97.6 |
| PZT-Radial[8] | 0.117 | 561 | **Yes** | 6.51 | 0.113 | 0.17 | 68 |
| PZT-Thickness[12] | 0.45 | - | **Yes** | 1 | 2.5 | 3.97 | 60 |
| **LN-SAW (This work)** | **39.35** | 715 | **Yes (2.1 kV)** | 1.66 | **8.47** | 17.5 †, 651‡ | 52.7 |

† Calculated using chip dimensions of 1.9 (long) × 1 (wide) × 0.05 (thick) cm$^3$. ‡ Calculated based on acoustic mode volume of 0.51 (long)×1 (wide)×0.005 (thick) cm$^3$.

REFERENCES

[1] Y. Zhang, D. Dong, Q. Li, R. Zhang, F. Udrea, and H. Wang, "Wide-bandgap semiconductors and power electronics as pathways to carbon neutrality," *Nature Reviews Electrical Engineering*, vol. 2, no. 3, pp. 155–172, Mar. 2025. DOI: https://doi.org/10.1038/s44287-024-00135-5

[2] Z. Yang, X. Yang, H. Gong, H. Cui, H. Zhang, N. Yan, X. Yu, Y. Qin, Y. Wang, X. Wang, C. Jiang, H. Wang, D. Dong, Q. Li, and Y. Zhang, "1000 V, 10 MHz Voltage Multiplier Based on Monolithically Integrated GaN-on-Sapphire Diode Bridge IC," *IEEE Transactions on Power Electronics*, vol. 41, no. 4, pp. 4578–4582, Apr. 2026. DOI: https://doi.org/10.1109/TPEL.2025.3642712

[3] Z. Yao, C. Daniel, E. Stolt, V. Chulukhadze, J. R. Davila, and R. Lu, "Lithium Tantalate Bulk Acoustic Resonator for Piezoelectric Power Conversion," in *2025 23rd International Conference on Solid-State Sensors, Actuators and Microsystems (Transducers)*, Jun. 2025, pp. 1882–1885, DOI: https://doi.org/10.1109/Transducers61432.2025.11110984.

[4] M. E. Galanko, S. S. Bedair, R. Rudy, V. F. G. Tseng, J. S. Pulskamp, and I. Kierzewski, "AT-Cut Quartz Piezoelectric Transformers for Passive Voltage Gain in an RF Front-End," *IEEE Electron Device Letters*, vol. 40, no. 10, pp. 1670–1673, Oct. 2019. DOI: https://doi.org/10.1109/LED.2019.2934053

[5] F. Qian, S. Chen, W. Wei, J. Xu, K. Yang, J. Zheng, Z. Ren, X. Liu, and Y. Yang, "Suppressing acoustomigration and temperature rise for high-power robust acoustics," *Nature Communications*, vol. 17, no. 1, pp. 5543, Apr. 2026. DOI: https://doi.org/10.1038/s41467-026-72102-7

[6] Z. Yao, C. Daniel, L. Matto, H. Chang, V. Chulukhadze, M. Liao, J. Kramer, E. Stolt, T. H. Hsu, M. S. Goorsky, J. Rivas-Davila, and R. Lu, "Periodically Poled Piezoelectric Lithium Niobate Resonator for Piezoelectric Power Conversion," *IEEE Transactions on Ultrasonics*, vol. 73, no. 7, pp. 861–870, Jul. 2026. DOI: https://doi.org/10.1109/TUSON.2026.3696612

[7] R. L. Lin, F. C. Lee, E. M. Baker, and D. Y. Chen, "Inductor-less piezoelectric transformer electronic ballast for linear fluorescent lamp," in *APEC 2001. Sixteenth Annual IEEE Applied Power Electronics Conference and Exposition (Cat. No. 01CH37181)*, Mar. 2001, vol. 2: IEEE, pp. 664–669, DOI: https://doi.org/10.1109/APEC.2001.912440.

[8] M. S. Roedgaard, M. Weirich, and M. A. E. Andersen, "Forward Conduction Mode Controlled Piezoelectric Transformer-Based PFC LED Drive," *IEEE Transactions on Power Electronics*, vol. 28, no. 10, pp. 4841–4849, Oct. 2013. DOI: https://doi.org/10.1109/tpel.2012.2233499

[9] C. Daniel, E. Stolt, W. Braun, R. Lu, and J. Rivas-Davila, "Nonlinear Losses and Material Limits of Piezoelectric Resonators for DC-DC Converters," presented at the *2024 IEEE Applied Power Electronics Conference and Exposition (APEC)*, 2024. DOI: https://doi.org/10.1109/apec48139.2024.10509514.

[10] S. Naval, W. Xu, M. Touhami, and J. D. Boles, "High-Efficiency Isolated Piezoelectric Transformers for Magnetic-less DC-DC Power Conversion," presented at the *2025 IEEE Applied Power Electronics Conference and Exposition (APEC)*, Jul., 2025. DOI: https://doi.org/10.1109/apec48143.2025.10977397.

[11] M. A. Tawfik, A. Ahmed, and J.-H. Park, "The Magnetic-Less Single-Stage Isolated DC/AC Converter Based on Piezoelectric Transformer," *IEEE Transactions on Circuits and Systems II: Express Briefs*, vol. 73, no. 3, pp. 338–342, Mar. 2026. DOI: https://doi.org/10.1109/tcsii.2026.3654066

[12] M. Sanz, P. Alou, A. Soto, R. Prieto, J. Cobos, and J. Uceda, "Magnetic-less converter based on piezoelectric transformers for step-down dc/dc and low power application," in *Eighteenth Annual IEEE Applied Power Electronics Conference and Exposition, 2003. APEC'03.*, Feb. 2003, vol. 2: IEEE, pp. 615–621, DOI: https://doi.org/10.1109/APEC.2003.1179278.

[13] E. Stolt, M. Affolter, Z. Ye, and J. Rivas-Davila, "Current mode control for high frequency piezoelectric resonator-based dc–dc converters," *IEEE Transactions on Power Electronics*, vol. 40, no. 9, pp. 12730–12738, Sep. 2025. DOI: https://doi.org/10.1109/TPEL.2025.3571873

[14] Z. Yao, H. Chang, E. Stolt, C. Daniel, T. H. Hsu, J. Rivas-Davila, and R. Lu, "Radial Mode Lithium Niobate Rosen Transformer," in *2026 IEEE 39th International Conference on Micro Electro Mechanical Systems (MEMS)*, Jan. 2026, pp. 1383–1386, DOI: https://doi.org/10.1109/MEMS64181.2026.11419439.

[15] J. D. Boles, J. E. Bonavia, J. H. Lang, and D. J. Perreault, "A Piezoelectric-Resonator-Based DC–DC Converter Demonstrating 1 kW/cm Resonator Power Density," *IEEE Transactions on Power Electronics*, vol. 38, no. 3, pp. 2811–2815, Mar. 2023. DOI: https://doi.org/10.1109/TPEL.2022.3217773

[16] W. D. Braun, E. A. Stolt, L. Gu, J. Segovia-Fernandez, S. Chakraborty, R. Lu, and J. M. Rivas-Davila, "Optimized Resonators for Piezoelectric Power Conversion," *IEEE Open Journal of Power Electronics*, vol. 2, no. pp. 212–224, Mar. 2021. DOI: https://doi.org/10.1109/ojpel.2021.3067020

[17] K. Nguyen, V. Chulukhadze, E. Stolt, W. Braun, J. Segovia-Fernandez, S. Chakraborty, J. Rivas, and R. Lu, "Near Spurious-Free Thickness Shear Mode Lithium Niobate Resonator for Piezoelectric Power Conversion," *IEEE Trans Ultrason Ferroelectr Freq Control*, vol. 70, no. 11, pp. 1536–1543, Nov. 2023. DOI: https://doi.org/10.1109/TUFFC.2023.3303123

[18] E. Stolt, W. Braun, K. Nguyen, V. Chulukhadze, R. Lu, and J. Rivas-Davila, "A Spurious-Free Piezoelectric Resonator Based 3.2 kW DC–DC Converter for EV On-Board Chargers," *IEEE Transactions on Power Electronics*, vol. 39, no. 2, pp. 2478–2488, Feb. 2024. DOI: https://doi.org/10.1109/tpel.2023.3334211

[19] M. Touhami, G. Despesse, and F. Costa, "A new topology of DC–DC converter based on piezoelectric resonator," *IEEE Transactions on Power Electronics*, vol. 37, no. 6, pp. 6986–7000, Jun. 2022. DOI: https://doi.org/10.1109/TPEL.2022.3142997

[20] Z. Cao, L. Shao, and L. Zhu, "Motional-Current-Sensing Method and Simplified Closed-Loop Control Strategy for Piezoelectric-Resonator-Based DC-DC Converters," *IEEE Transactions on Power Electronics*, vol. no. pp. 1–14, Nov. 2026. DOI: https://doi.org/10.1109/TPEL.2026.3703942

[21] C. Campbell, *Surface acoustic wave devices and their signal processing applications*. Elsevier, 2012.

[22] L. Shao, D. Zhu, M. Colangelo, D. Lee, N. Sinclair, Y. Hu, P. T. Rakich, K. Lai, K. K. Berggren, and M. Lončar, "Electrical control of surface acoustic waves," *Nature Electronics*, vol. 5, no. 6, pp. 348–355, Jun. 2022. DOI: https://doi.org/10.1038/s41928-022-00773-3

[23] Z. Xi, Z. Cen, D. Wang, J. G. Thomas, B. R. Srijanto, I. I. Kravchenko, J. Zuo, H. Liu, J. Ji, and Y. Zhu, "Room-Temperature Mid-Infrared Detection Using Metasurface-Absorber-Integrated Phononic Crystal Oscillator," *Laser & Photonics Reviews*, vol. 19, no. 20, pp. e00498, Jul. 2025. DOI: https://doi.org/10.1002/lpor.202500498

[24] L. Jin, Z. Xi, J. G. Thomas, J. Ji, Y. Zhang, N. Chen, Y. Zhu, L. Shao, and L. Zhu, "Microwave-acoustic-based isolated gate driver for power electronics," *Communications Engineering*, vol. 5, no. 1, pp. 125, May. 2026. DOI: https://doi.org/10.1038/s44172-026-00681-w

[25] H. Tokuyama, K. Wada, and S. Goka, "Design and Experiment of a Gate Drive Circuit using SAW Filter with Signal and Power Transmission," in *2019 IEEE 4th International Future Energy Electronics Conference (IFEEC)*, Nov. 2019: IEEE, pp. 1–4, DOI: https://doi.org/10.1109/IFEEC47410.2019.9014985.

[26] M. Chaudhari, L. Matto, N. Ahmed, M. Liao, V. Tallavajhula, Y. Long, Z. Yao, J. Campbell, T. H. Hsu, M. S. Goorsky, and R. Lu, "Thermal Endurance of Suspended Thin-Film Lithium Niobate up to 800 ∘C," *IEEE Transactions on Ultrasonics*, vol. 73, no. 3, pp. 315–325, Mar. 2026. DOI: https://doi.org/10.1109/TUSON.2025.3650497

[27] J. G. Thomas, Z. Xi, J. Ji, G. Shi, B. R. Srijanto, I. I. Kravchenko, Y. Yao, L. Shao, and Y. Zhu, "Spectral interferometry-based microwave-frequency vibrometry for integrated acoustic wave devices," *Optica*, vol. 12, no. 7, pp. 935–944, Jul. 2025. DOI: https://doi.org/10.1364/OPTICA.560740

[28] K. S. Van Dyke, "The piezo-electric resonator and its equivalent network," *Proceedings of the Institute of Radio Engineers*, vol. 16, no. 6, pp. 742–764, Jun. 1928. DOI: https://doi.org/10.1109/JRPROC.1928.221466

[29] D. M. Pozar, *Microwave engineering*. John wiley & sons, 2011.

[30] A. M. Flynn and S. R. Sanders, "Fundamental limits on energy transfer and circuit considerations for piezoelectric transformers," *IEEE transactions on power electronics*, vol. 17, no. 1, pp. 8–14, Jan. 2002. DOI: https://doi.org/10.1109/63.988662

[31] E. A. Stolt, C. Daniel, and J. M. Rivas-Davila, "A stacked radial mode lithium niobate transformer for DC-DC conversion," in *2024 IEEE Workshop on Control and Modeling for Power Electronics (COMPEL)*, Jun. 2024: IEEE, pp. 1–7, DOI: https://doi.org/10.1109/COMPEL57542.2024.10613959.